# A Robust and Distribution-Fitting-Free Estimation Approach of Travel Time Percentile Function based on L-moments

Ruiya Chen, Xiangdong Xu, Jianqiang Li

*Abstract*— Travel time is one of the key indicators monitored by intelligent transportation systems, helping the systems to gain real-time insights into traffic situations, predict congestion, and identify network bottlenecks. Travel time exhibits variability, and thus suitable probability distributions are necessary to accurately capture full information of travel time variability. Considering the potential issues of insufficient sample size and the disturbance of outliers in actual observations, as well as the heterogeneity of travel time distributions, we propose a robust and distribution-fitting-free estimation approach of travel time percentile function using L-moments based Normal-Polynomial Transformation. We examine the proposed approach from perspectives of validity, robustness, and stability based on both theoretical probability distributions and real data. The results indicate that the proposed approach exhibits high estimation validity, accuracy and low volatility in dealing with outliers, even in scenarios with small sample sizes.

## I. Introduction

Accurately monitoring traffic situations is one of the main tasks of intelligent transportation systems (ITS). Through dynamic monitoring of traffic situations, ITS can predict traffic peak periods, identify network bottleneck, and implement adaptive traffic control strategies to ensure the healthy operation of transportation systems. Travel time is an important indicator of evaluating the operational status of transportation systems, and accurately characterizing travel time is the basis for understanding traffic situations [1], [2], [3]. However, due to the complex interactions among travelers, vehicles, roads, and the environment, travel time exhibits variability, and many empirical studies have found that it is often characterized by right-skewness and long tails [4], [5]. Therefore, appropriate probability distributions are needed to represent the complete information of travel time variability. The travel time distribution also serves as the data foundation for analysis of travel time reliability.

Many probability distributions have been used to characterize travel time variability, such as Normal [6], Lognormal [7], Weibull [8], Gamma [9], Burr [10], (Generalized) Extreme Value distributions [11], etc., for details please refer to the review by Zang et al [12]. However, travel time distribution exhibits heterogeneity, with its characteristics varying across different links/paths and time periods. Currently, there is no consensus on the selection of the optimal fitted distribution. Additionally, the choice of the optimal fitted distribution by different goodness-of-fit tests (such as K-S test, $\chi^2$ test, and $R^2$) may also vary. Zang et al. proposed a distribution-fitting-free method to estimate the Percentile Function of Travel time (PTT) based on Cornish-Fisher expansion (CF) [13]. CF demonstrates greater adaptability to diverse travel time distributions with lower estimation error compared to the aforementioned probability distributions.

Travel time datasets inevitably contain outliers. Although filtering algorithms can be used for data cleaning, each algorithm has different filtering criteria and limitations, thus it cannot guarantee the complete removal of outliers from the dataset. CF requires the central moments of travel time distributions as inputs, and central moments are overly sensitive to outliers, especially for higher-order central moments (such as skewness and kurtosis which are utilized in CF) [14]. The estimated PTT lack robustness and may lead planners to biased or even erroneous decisions. Furthermore, under limited sample sizes, sample skewness and kurtosis are bounded, and the size of the boundary values depends on the sample size [15]. For highly-skewed and long-tailed travel time distributions, the sample skewness and kurtosis are difficult to accurately represent the true population skewness and kurtosis when the sample size is very limited. Some scholars have found that even under large sample sizes, the estimation errors of traditional skewness and kurtosis are still significant [16]. In practical observations, from the perspective of the entire road network, the sample size of most links/paths is relatively limited, and this limitation restricts the applicability of CF in estimating PTT.

Considering the observations mentioned above, we propose a distribution-fitting-free PTT estimation approach based on L-moments rather than central moments. L-moments possess unbiased estimators, are capable of effectively reducing the influence of outliers, and have lower requirements for sample size. In our previous study, we theoretically and empirically verified the superiority of L-skewness and L-kurtosis over traditional skewness and kurtosis (used by CF) [14].

The remainder of this paper is as follows: Section 2 introduces the proposed approach, including Normal-Polynomial Transformation (NPT), L-moments, parameter solving method of L-moments based NPT, and the validity domain of L-moments based NPT; in Section 3, we examine the robustness, validity, and stability of L-moments based NPT and CF using commonly used travel time probability distributions and real travel time datasets as benchmarks; Section 4 summarizes the paper.

Ruiya Chen is with the School of Transportation Engineering, Kunming University of Science and Technology, Kunming 650550, China (e-mail: chenruiya@kust.edu.cn).

Xiangdong Xu is the Key Laboratory of Road and Traffic Engineering, Tongji University, Shanghai, 201804 China (e-mail: xiangdongxu@tongji.edu.cn).

Jianqiang Li is the Key Laboratory of Road and Traffic Engineering, Tongji University, Shanghai 201804, China (e-mail: ljq19920202@163.com).

The corresponding author is Xiangdong Xu.

## II. METHODOLOGY

In this section, we introduce the estimation approach for PTT by using the L-moments based NPT (LMNPT).

### A. Normal-Polynomial Transformation (NPT)

Fleishman proposed NPT to express nonnormal random variables via a third-order polynomial of standard normal random variable [17]. Following the NPT, we can approximate PTT as shown in (1):

$$PTT(p) = a + b\Phi^{-1}(p) + c\left[\Phi^{-1}(p)\right]^2 + d\left[\Phi^{-1}(p)\right]^3 \quad (1)$$

where $PTT(p)$ denotes the percentile function of travel time; $a$, $b$, $c$, and $d$ are coefficients of NPT; and $\Phi^{-1}(p)$ is the inverse standard normal cumulative distribution function (CDF).

The fundamental idea of determining the coefficients in (1) is to link the first four moments (e.g., by using raw moments [17]) of the original random variable with those of the standard normal variable. Specifically, make the first four moments of travel time equal to the first four moments of the right-hand side of (1), and we can have four equations. By solving this system of equations simultaneously, we can determine the coefficients. In addition, CF utilizes Hermite polynomial to adjust the standard normal variable to the nonnormal variable [18]. Zang et al. utilized the CF which employed the first four central moments (mean, variance, skewness, and excess kurtosis) for estimating the PTT [13]. However, the raw moments and central moments are both sensitive to the outliers, especially for higher-order moments. Besides, they are limited by sample size especially for higher-order moments, and the required sample size for estimation can be very large, which may not be satisfied in practice. The above two disadvantages may reduce the accuracy of the estimated PTT by using raw moments and central moments. In this paper, we adopt more robust L-moments to approximate PTT. Below we give a brief introduction of L-moments.

### B. L-moments

Hosking proposed moments based on order statistics [19], which can be expressed as a linear combination of expected order statistics, thus termed as linear moments (L-moments). The formula for L-moments is:

$$l_r = r^{-1} \sum_{k=0}^{r-1}(-1)^k \binom{r-1}{k} E(T_{r-k:r}), \quad r = 1, 2, 3\cdots \quad (2)$$

where $l_r$ represents $r^{th}$-order L-moments; $T_{r-k:r}$ represents the order statistic of the travel time random variable $T$, when the sample size is $r$ and the random variables are arranged in ascending order, with the order being $r - k$. The expectation of the order statistic with order $j$ and sample size $r$ (i.e., $E[T_{j:r}]$) can be expressed as:

$$E(T_{j:r}) = \frac{r!}{(j-1)!(r-j)!} \int t\left[F(t)\right]^{j-1}\left[1-F(t)\right]^{r-j} dF(t)$$

$$j = 1, 2, \cdots, r \quad (3)$$

where $F(t)$ is CDF of travel time. According to (2) and (3), the first four L-moments can be expressed by $PTT(p)$ as follows:

$$l_1 = E(T) = \int_0^1 PTT(p)dp$$
$$l_2 = \frac{1}{2}E(T_{2:2} - T_{1:2}) = \int_0^1 PTT(p)(2p-1)dp$$
$$l_3 = \frac{1}{3}E(T_{3:3} - 2T_{2:3} + T_{1:3}) = \int_0^1 PTT(p)(6p^2 - 6p + 1)dp \quad (4)$$
$$l_4 = \frac{1}{4}E(T_{4:4} - 3T_{3:4} + 3T_{2:4} - T_{1:4})$$
$$= \int_0^1 PTT(p)(20p^3 - 30p^2 + 12p - 1)dp$$

In order to compute sample L-moments, it is necessary to iterate through all sub-samples of size $r$, and then calculate the mean of the observed travel time with order $j$ within each sub-sample. To simplify this process and avoid such iteration, probability weighted moments (PWM) can be utilized, which can also be expressed by $PTT(p)$:

$$\beta_q = E(p^q PTT(p)), \quad q = 0, 1, 2, \cdots \quad (5)$$

where $\beta_q$ denotes $q^{th}$-order PWM. According to (4) and (5), we can rewrite the first four L-moments as follows:

$$\begin{aligned}l_1 &= \beta_0 \\ l_2 &= 2\beta_1 - \beta_0 \\ l_3 &= 6\beta_2 - 6\beta_1 + \beta_0 \\ l_4 &= 20\beta_3 - 30\beta_2 + 12\beta_1 - \beta_0\end{aligned} \quad (6)$$

Thus, once $\beta_q$ is estimated, the sample L-moments can be obtained through (6). The unbiased estimation method for $\beta_q$ can be referred to the research by Landwehr et al [20]. Due to the unbiasedness, approximate normality, and robustness to outliers of sample order statistics, sample L-moments also inherit these advantages.

Similar to the skewness and kurtosis based on central moments, L-skewness ($\tau_3$) and L-kurtosis ($\tau_4$) is defined as:

$$\begin{aligned}\tau_3 &= l_3 / l_2 \\ \tau_4 &= l_4 / l_2\end{aligned} \quad (7)$$

Chen et al. proposed using L-skewness and L-kurtosis to summarize travel time distributions rather than skewness and kurtosis, considering that they perform better in unbiasedness, robustness and effectiveness [14].

### C. Approximating Travel Time Percentile Function by L-moments based NPT

As shown in (6), the first four L-moments can be expressed by $\beta_q$, and with (1) we rewritten the first four $\beta_q$ as follows:

$$\begin{aligned}\beta_q &= \int_0^1 p^q PTT(p) \\ &= \int_0^1 p^q \left[a + b\Phi^{-1}(p) + c\left[\Phi^{-1}(p)\right]^2 + d\left[\Phi^{-1}(p)\right]^3\right]dp \\ &= \int_0^1 \left[ap^q + bp^q\Phi^{-1}(p) + cp^q\left[\Phi^{-1}(p)\right]^2 + dp^q\left[\Phi^{-1}(p)\right]^3\right]dp \\ &= aS_{q,0} + bS_{q,1} + cS_{q,2} + dbS_{q,3}\end{aligned}$$

$$(8)$$

where $S_{q,m}$ is defined as:

$$S_{q,m} = \int_0^1 p^q \left[\Phi^{-1}(p)\right]^m dp \quad m = 0, 1, 2, 3 \quad (9)$$

Hence, employing the numerical values of $S_{q,m}$, computed by Tung [21], enables an explicit and straightforward solution for the NPT coefficients. The coefficients in (1) can thus be calculated by the first four L-moments as follows:

$$\begin{aligned} a &= l_1 + a_1 l_3 \\ b &= b_1 l_2 + b_2 l_4 \\ c &= c_1 l_3 \\ d &= d_1 l_2 + d_2 l_4 \end{aligned} \quad (10)$$

where $a_1 = -1.81379937$, $b_1 = 2.25518617$, $b_2 = -3.9374025$, $c_1 = -a_1$, $d_1 = -0.19309293$, and $d_2 = 1.574961$.

### D. Validity Domain of L-moments based NPT

To ensure a monotonic estimated PTT, below we derive the validity domain of LMNPT. The monotone estimated PTT means that $dPTT(p)/dp$ should be nonnegative:

$$\frac{dPTT(p)}{dp} = \frac{d\Phi^{-1}(p)}{dp}\left[b + 2c\Phi^{-1}(p) + 3d\left[\Phi^{-1}(p)\right]^2\right] \geq 0 \quad (11)$$

Given that $\Phi^{-1}(p)$ is monotone nondecreasing, the first term on the right-hand side of (11) is nonnegative. Then, we rewrite the term in the outer square brackets as follows:

$$g\left[\Phi^{-1}(p)\right] = b_1 l_2 + b_2 l_4 + 2 c_1 l_3 \Phi^{-1}(p) + 3(d_1 l_2 + d_2 l_4)\left[\Phi^{-1}(p)\right]^2 \quad (12)$$

(12) should be nonnegative to ensure the monotonicity of PTT. Given that $l_2 > 0$, (12) can be simplified as:

$$g\left[\Phi^{-1}(p)\right] = b_1 + b_2 \tau_4 + 2 c_1 \tau_3 \Phi^{-1}(p) + 3(d_1 + d_2 \tau_4)\left[\Phi^{-1}(p)\right]^2 \geq 0 \quad (13)$$

There are two cases that can satisfy (13): (a) $g[\Phi^{-1}(p)]$ is equal to a positive constant value, we have:

$$\begin{cases} 3(d_1 + d_2 \tau_4) = 0 \\ \tau_3 = 0 \\ b_1 + b_2 \tau_4 \geq 0 \end{cases} \quad (14)$$

and (b) $g[\Phi^{-1}(p)]$ is a quadratic function with respect to $\Phi^{-1}(p)$. For the second case, the coefficient of the quadratic term must be positive and the discriminant must be non-positive, and thus we have:

$$\begin{cases} 3(d_1 + d_2 \tau_4) > 0 \\ \Delta_g = h(\tau_3) = c_1^2 \tau_3^2 - 3(d_1 + d_2 \tau_4)(b_1 + b_2 \tau_4) \leq 0 \end{cases} \quad (15)$$

where $\Delta_g$ is the discriminant of $g[\Phi^{-1}(p)]$. Similarly, $h(\tau_3)$ is also a quadratic function. Given that $c_1^2 > 0$, $h(\tau_3) \leq 0$ means that $h(\tau_3) = 0$ has two roots and $\Delta_g \leq 0$ when $\tau_3$ is between the two roots. Then we have:

$$(d_1 + d_2 \tau_4)(b_1 + b_2 \tau_4) < 0 \quad (16)$$

The validity domain of the LMNPT is (15)∩(16)∪(14) as shown in (17), which is visualized in Figure 1.

$$\begin{cases} -\dfrac{d_1}{d_2} \leq \tau_4 \leq -\dfrac{b_1}{b_2} \\ \Delta_g = h(\tau_3) = c_1^2 \tau_3^2 - 3(d_1 + d_2 \tau_4)(b_1 + b_2 \tau_4) \leq 0 \end{cases} \quad (17)$$

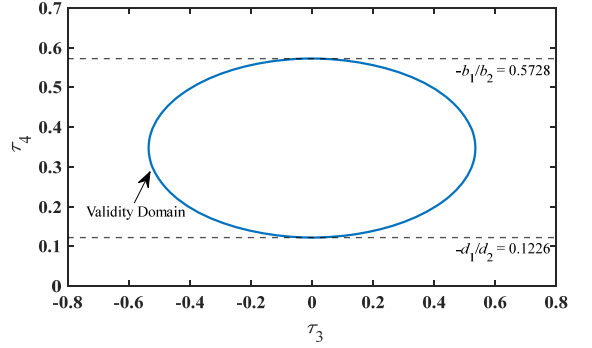

Figure 1. The validity domain of the L-moments based NPT for estimating PTT.

## III. PERFORMANCE VALIDATION

This section explores the superiority of LMNPT in estimating PTT compared to CF which has been shown to perform well in approximating PTT [13]. We conduct the performance validation in terms of the validity, robustness, and stability, using both the widely-used travel time probability distributions (i.e., Normal, Burr, Extreme Value, Gamma, Lognormal, and Weibull distributions) with different coefficient of variation (CoV) and the realistic travel time datasets as the true values of travel times. Note that we utilize the improved CF proposed by Zang et al [13].

### A. Performance for Estimating Travel Time Probability Distribution

To explicitly explore the effects of statistical characteristics, sampling, sample size on estimation performance of LM, we adopt theoretical probability distributions as the true values of travel times. The specific validation methods are as follows:

- The six widely-used travel time probability distributions are divided in three groups by CoV (i.e., CoV = 0.07, 0.15, and 0.30), and we examine the estimation performance of LM and CF, respectively. Note that the six distributions maintain the same mean value of 167.

- Random sampling is conducted for each distribution, with a sample size of 100 for each sampling, repeated 100 times. Based on the estimation of the above 100 trials, we analyze the two estimation methods in terms of the average estimation performance and the volatility of estimation performance.

- The sample size gradually increases from 100 to 2000 in increments of 100 to explore the impact of sample size on estimation performance.

- To test the robustness to outliers, outliers are artificially added to each set of sampled data. The data with added outliers is then used as new samples to estimate the original distribution. Outliers are defined as:

$$t_{outlier} = 1.5 t_{max} \text{ or } 0.5 t_{min} \quad (18)$$

where $t_{outlier}$ is the added outlier, and $t_{max}$ and $t_{min}$ are the upper bound and lower bound of the travel time sample set.

- We assess the estimation performance by validity rate (VR), $\chi^2$, mean absolute percentage error (MAPE), root mean square error (RMSE), and $R^2$.

*1) Normal and Burr with CoV = 0.07*

Figure 2. shows the estimated PTTs under different scenarios by the two methods based on 100 samples. We can see that when adding outliers, the Cornish-Fisher expansion shows a poor performance of estimating two distributions, especially for adding the outlier of $0.5t_{min}$. However, the proposed LMNPT maintains good estimation performance under different scenarios.

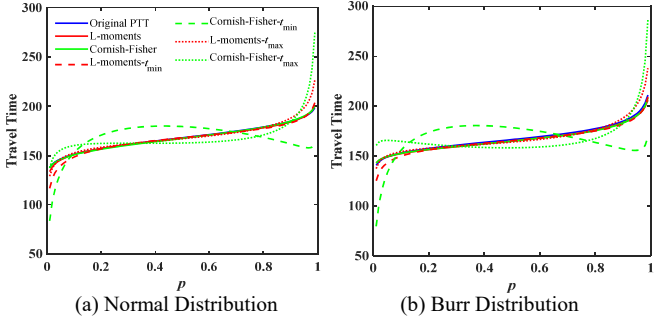

Figure 2. The estimated PTTs for Normal and Burr distributions (*Note: L-moments means the L-moments based NPT; L-moments-$t_{max}$ means the manipulation of adding the outlier with $1.5t_{max}$, and so on*)

TABLE I. shows the average estimation performance and the volatility of estimation performance for Normal distribution under random sampling. We can see that when outliers are not added, the performance of the two methods is comparable. When adding the outlier with $0.5t_{min}$, CF fails to estimate the Normal distribution since the validity rate is zero; on the contrary, the proposed LMNPT achieves a validity rate of 100% while maintaining high accuracy. Additionally, the proposed method exhibits lower volatility, as indicated by lower standard deviations of performance metrics. The performance under Burr distribution is similar to that under Normal distribution; therefore, it will not be further elaborated here.

TABLE I. ESTIMATION PERFORMANCE UNDER DIFFERENT OUTLIERS FOR NORMAL DISTRIBUTION

| Method | Scenario | VR | $\chi^2$ | MAPE | RMSE | $R^2$ |
|---|---|---|---|---|---|---|
| LM | without adding | 100% (0) | 1.73 (1.64) | 0.78% (0.42%) | 1.55 (0.69) | 0.98 (0.02) |
|  | $0.5t_{min}$ | 100% (0) | 6.78 (3.17) | 1.25% (0.42%) | 3.11 (0.74) | 0.95 (0.02) |
|  | $1.5t_{max}$ | 100% (0) | 9.88 (3.65) | 1.34% (0.37%) | 4.25 (0.76) | 0.92 (0.03) |
| CF | without adding | 100% (0) | 1.63 (1.63) | 0.77% (0.42%) | 1.49 (0.71) | 0.98 (0.02) |
|  | $0.5t_{min}$ | 0 (0) | 181.23 (26.14) | 9.12% (0.64%) | 17.56 (1.39) | -0.12 (0.11) |
|  | $1.5t_{max}$ | 34% (47.61%) | 72.08 (13.45) | 4.19% (0.56%) | 11.54 (1.05) | 0.53 (0.08) |

a. values in parentheses is standard deviation.

Figure 3. shows the estimation performance under different sample size for Normal distribution. As the sample size increases, the estimation performance of CF improves more than LM does. However, even when the sample size increases to 2000, its performance still lags behind the proposed LMNPT. The estimation performance of the proposed method does not require a high sample size; even with a small sample size, it can still maintain good estimation performance. Therefore, in practical applications, it can be applicable even when there are not a large number of travel time data samples available for analysis within the time window. The results for Burr distribution are consistent with those for Normal distribution, and thus not presented herein.

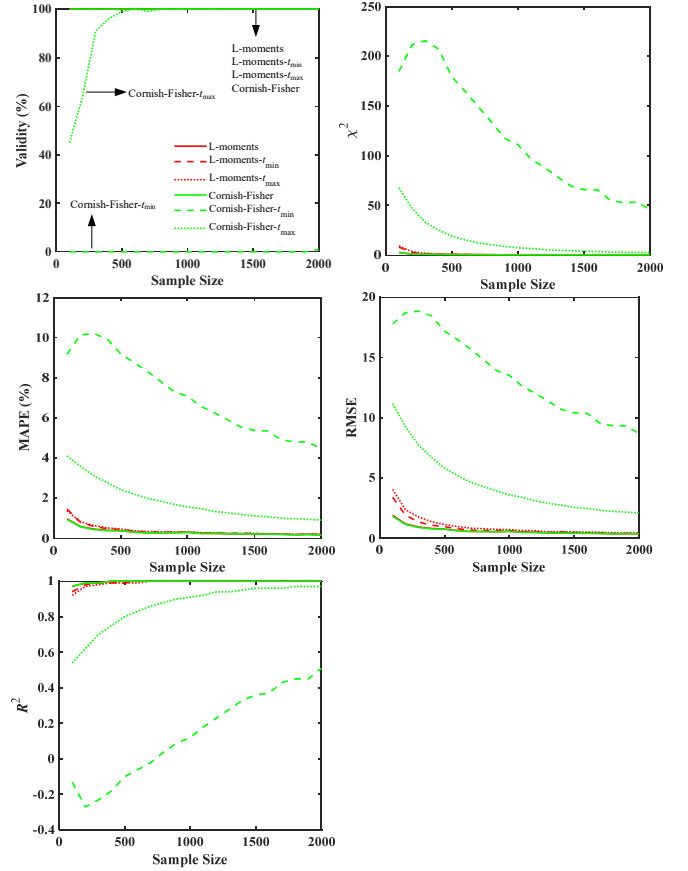

Figure 3. The impact of sample size on estimation performance for Normal distribution

*2) Extreme Value and Gamma with CoV = 0.15*

Figure 4. shows the estimated PTTs under different scenarios by the two methods based on 100 samples. TABLE II. shows the average estimation performance and the volatility of estimation performance for Extreme Value distribution under random sampling. We can see that without adding outliers under the Extreme Value distribution, the validity rate of CF is 36%, and the validity rate of LMNPT is 96%. Even without adding outliers, the estimation performance of CF is far inferior to LMNPT. When adding outliers, although the performance of CF is better under the Extreme Value distribution compared to the Normal and Burr distributions, it still does not outperform LM. The performance under the Gamma distribution is similar to that under Normal and Burr distributions.

Figure 5. shows the estimation performance under different sample size for Extreme Value distribution. We can observe significant volatility in the performance of CF as the sample size varied. This could be attributed to the generation of some extreme values during the sampling process. Due to CF's sensitivity to extreme values, it results in larger fluctuations. In contrast, LMNPT exhibits lower volatility while maintaining good estimation validity and accuracy. The findings regarding Gamma distribution align with those of Normal and Burr distributions.

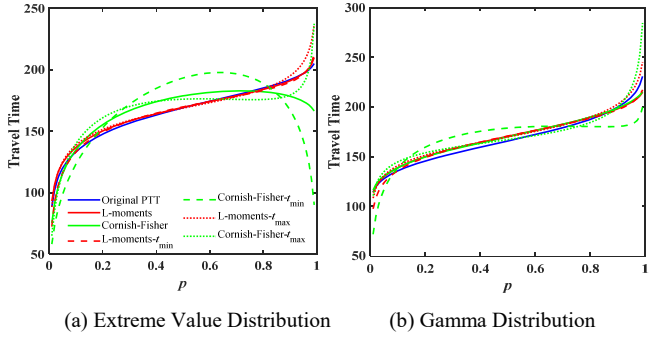

(a) Extreme Value Distribution    (b) Gamma Distribution

Figure 4. The estimated PTTs for Extreme Value and Gamma Distributions

TABLE II. ESTIMATION PERFORMANCE UNDER DIFFERENT OUTLIERS FOR EXTREME VALUE DISTRIBUTION

| Method | Scenario | VR | $\chi^2$ | MAPE | RMSE | $R^2$ |
|---|---|---|---|---|---|---|
| LM | without adding | 96% (19.69%) | 10.23 (8.42) | 1.76% (0.82%) | 3.40 (1.39) | 0.97 (0.03) |
| | $0.5t_{min}$ | 99% (10.00%) | 17.25 (15.88) | 2.20% (1.11%) | 4.15 (1.96) | 0.97 (0.02) |
| | $1.5t_{max}$ | 100% (0) | 17.56 (8.65) | 2.19% (0.75%) | 5.19 (1.10) | 0.95 (0.03) |
| CF | without adding | 36% (48.24%) | 52.85 (155.12) | 3.40% (4.20%) | 6.55 (6.95) | 0.89 (0.21) |
| | $0.5t_{min}$ | 2% (14.07%) | 207.20 (293.87) | 9.27% (6.89%) | 16.81 (8.88) | 0.63 (0.18) |
| | $1.5t_{max}$ | 90% (30.15%) | 62.43 (141.18) | 3.74% (3.72%) | 8.52 (5.93) | 0.86 (0.19) |

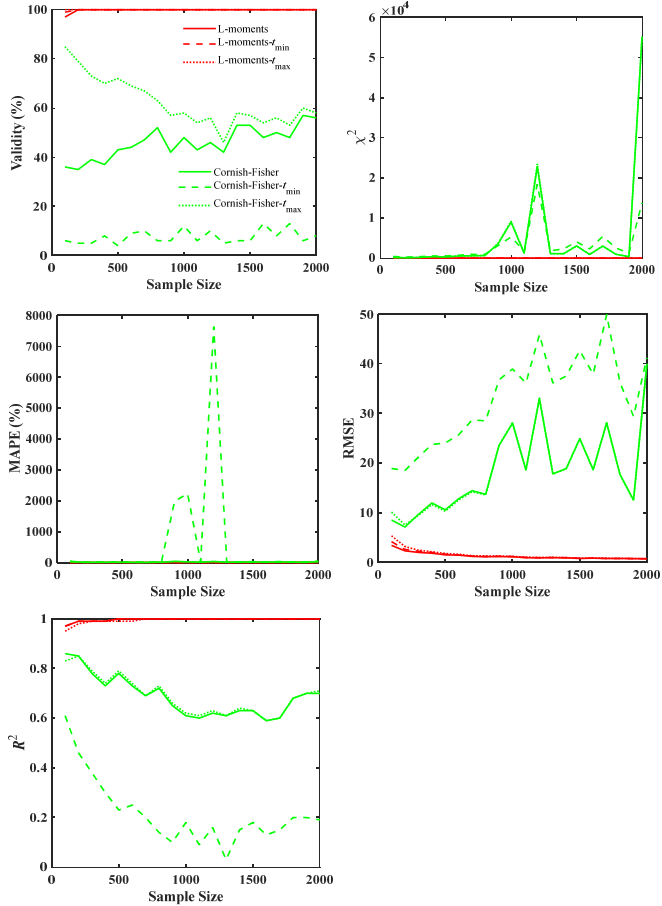

Figure 5. The impact of sample size on estimation performance for Extreme Value distribution

### 3) Lognormal and Weibull with CoV = 0.30

Figure 6. shows the estimated PTTs under different scenarios by the two methods based on 100 samples. TABLE III. and TABLE IV. show the average estimation performance and the volatility of estimation performance under random sampling for Lognormal and Weibull distributions. We found that for Lognormal and Weibull distributions, there is no much difference in estimation performance between CF and LMNPT. When adding the outlier with $0.5t_{min}$, LMNPT performs slightly better than CF.

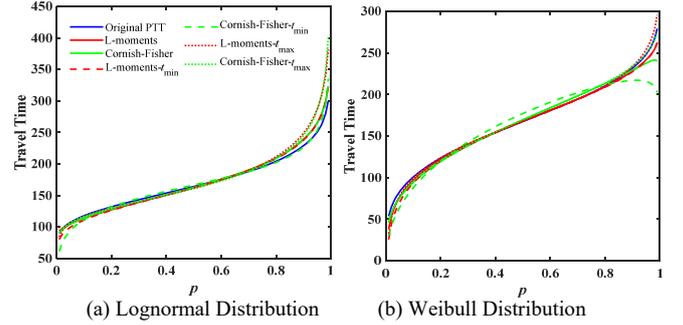

(a) Lognormal Distribution    (b) Weibull Distribution

Figure 6. The estimated PTTs for Lognormal and Weibull Distributions

Similar to Normal, Burr, and Gamma distributions, with increasing sample size, the estimation performance of CF improves more than that of LMNPT for these two distributions. Even when the sample size increases to 2000, CF's performance still falls short of the proposed LMNPT under the scenario of adding outliers. For brevity, we exclude the figures related to these two distributions.

TABLE III. ESTIMATION PERFORMANCE UNDER DIFFERENT OUTLIERS FOR LOGNORMAL DISTRIBUTION

| Method | Scenario | VR | $\chi^2$ | MAPE | RMSE | $R^2$ |
|---|---|---|---|---|---|---|
| LM | without adding | 99% (10%) | 21.15 (15.74) | 2.75% (1.27%) | 5.93 (2.39) | 0.98 (0.02) |
| | $0.5t_{min}$ | 100% (0) | 25.97 (18.37) | 3.13% (1.37%) | 6.38 (2.25) | 0.98 (0.02) |
| | $1.5t_{max}$ | 100% (0) | 49.67 (42.16) | 3.25% (1.26%) | 9.70 (4.74) | 0.96 (0.03) |
| CF | without adding | 100% (0) | 20.06 (15.18) | 2.67% (1.28%) | 5.80 (2.39) | 0.98 (0.02) |
| | $0.5t_{min}$ | 100% (0) | 62.54 (31.06) | 4.96% (1.45%) | 9.86 (2.81) | 0.93 (0.05) |
| | $1.5t_{max}$ | 100% (0) | 60.76 (52.59) | 3.31% (1.26%) | 11.14 (5.22) | 0.94 (0.04) |

TABLE IV. ESTIMATION PERFORMANCE UNDER DIFFERENT OUTLIERS FOR WEIBULL DISTRIBUTION

| Method | Scenario | VR | $\chi^2$ | MAPE | RMSE | $R^2$ |
|---|---|---|---|---|---|---|
| LM | without adding | 99.90% (0.45%) | 6.20 (7.96) | 1.42% (0.79%) | 2.41 (1.29) | 1 (0.01) |
| | $0.5t_{min}$ | 99.90% (0.45%) | 7.02 (9.96) | 1.51% (0.98%) | 2.46 (1.40) | 1 (0.01) |
| | $1.5t_{max}$ | 100% (0) | 7.85 (11.80) | 1.51% (0.91%) | 2.79 (1.86) | 1 (0.01) |
| CF | without adding | 97.60% (5.07%) | 12.93 (10.78) | 1.99% (0.79%) | 3.86 (1.32) | 0.99 (0.01) |
| | $0.5t_{min}$ | 96.30% (6.79%) | 50.72 (23.32) | 4.07% (1.08%) | 7.48 (1.88) | 0.95 (0.02) |
| | $1.5t_{max}$ | 99.55% (0.83%) | 12.74 (12.84) | 1.94% (0.83%) | 3.67 (1.71) | 0.99 (0.01) |

Overall, LMNPT performs comparably to CF when outliers are not added. However, when facing outliers, LMNPT's validity, accuracy, and stability make it outperform

CF. Since outliers are inevitable in the real observed travel time datasets, LMNPT is more suitable for real-world applications to provide reliable travel time information for analysis of travel time reliability.

### B. Performance for Estimating Empirical Travel Time

To further investigate the applicability of LMNPT in estimating the PPT in practice, we conduct the performance validation by using the realistic travel time datasets extracted from License Plate Recognition (LPR) system in Shenzhen, China. Detailed information regarding the data description can refer to Chen et al [22]. Note that the empirical PTTs serve as the true values of PTTs.

TABLE V. shows the average estimation performance and the volatility of estimation performance for empirical travel time under different scenarios. When outliers are not added, the estimation performance of LMNPT and CF is almost the same. However, after adding outliers, LMNPT still maintains a high validity rate (i.e., VR = 96.03%) while CF's validity rate drops from 97.35% to 13%. Meanwhile LM's estimation accuracy remains at a lower level (i.e., MAPE = 0.56%), while CF's MAPE increases from 0.54% to 7.51%. In addition, under the scenarios of adding outliers, LMNPT shows a lower volatility of estimation performance than CF does rendering smaller standard deviations of the metrics.

The results are consistent with those under the widely-used travel time probability distributions. The proposed LMNPT shows superiority over CF in terms of validity, robustness and stability in estimating PTTs.

TABLE V.  ESTIMATION PERFORMANCE FOR EMPRICAL TRAVEL TIMES

| Method | Scenario | VR | $\chi^2$ | MAPE | RMSE | $R^2$ |
|---|---|---|---|---|---|---|
| LM | without adding | 94.37% (23.09%) | 0.91 (1.26) | 0.41% (0.23%) | 1.22 (0.86) | 0.99 (0.01) |
|  | $0.5t_{min}$ | 96.03% (19.57%) | 2.18 (3.49) | 0.56% (0.38%) | 1.67 (1.19) | 0.99 (0.02) |
|  | $1.5t_{max}$ | 95.70% (20.33%) | 3.39 (5.99) | 0.55% (0.42%) | 2.16 (1.72) | 0.98 (0.03) |
| CF | without adding | 97.35% (16.09%) | 1.36 (2.03) | 0.54% (0.33%) | 1.46 (1.11) | 0.99 (0.01) |
|  | $0.5t_{min}$ | 13.58% (34.31%) | 155.96 (89.43) | 7.51% (2.40%) | 16.94 (6.60) | 0.24 (0.35) |
|  | $1.5t_{max}$ | 87.42% (33.22%) | 21.36 (17.93) | 2.11% (0.86%) | 6.41 (2.64) | 0.86 (0.11) |

## IV. CONCLUSION

Percentile Function of Travel time (PTT) is a common component of almost all travel time reliability measures in the literature. This paper proposes estimating the PTT by L-moments based NPT (LMNPT). The validity domain of LMNPT is theoretically demonstrated, and its superiority over central moments-based Cornish-Fisher expansion is verified on the aspects of validity, robustness, and stability based on both theoretical probability distributions and real datasets. Outliers unavoidably exist in the realistic travel time datasets, LMNPT can mitigate the impact of outliers, and provide reliable trustworthy travel time information to assist travel scheduling and analysis of transportation systems.